\documentclass[10pt,letterpaper]{article}

\usepackage{color}
\usepackage{opex3}
\usepackage{amsmath}
\usepackage{amssymb}
\usepackage{graphicx}
\usepackage[english]{babel}
\usepackage{multirow}
\usepackage{array}

\newcommand{\fref}[1]{Fig. \ref{#1}}
\newcommand{\eref}[1]{(\ref{#1})}
\newcommand{\tref}[1]{Tab. \ref{#1}}
\renewcommand{\vec}[1]{\mathbf{#1}}

\begin{document}

\title{Role of edge inclination in optical microdisk resonator for label-free sensing}

\author{D.~Gandolfi$^1\,^*$, F.~Ramiro~Manzano$^1$, F.~J.~Aparicio~Rebollo$^1$, M.~Ghulinyan$^2$, G.~Pucker$^2$, L.~Pavesi$^1$}
\address{$^1$Nanoscience Laboratory, Dept. Physics, University of Trento, Via Sommarive 14, I-38123 Trento, Italy}
\address{$^2$Centre for Materials and Microsystems, Fondazione
 Bruno Kessler, via Sommarive 18, I-38123 Trento, Italy}

\email{$^*$d.gandolfi@science.unitn.it}

\begin{abstract}
In this paper we report on the measurement and modelling of enhanced optical refractometric sensors based on whispering-gallery-modes. The devices under test are optical microresonators made of silicon nitride on silicon oxide. In our approach, these microresonators are vertically coupled to a buried waveguide with the aim of creating integrated and cost-effective devices. 
The optimization analysis is a delicate balance of resonance quality factor and evanescent field overlap with the sorrounding environment to analyze. By numerical simulations we show that the microdisk thickness is critical to yield high figure of merit for the sensor, while edge inclination is less important. We also show that figures of merit as high as $1600/\mathrm{RIU}$ are feasible.
\end{abstract}



\section*{Introduction}
It was more than a decade ago when a whispering-gallery-mode (WGM) optical resonator has been used to experimentally prove the feasibility of a label-free biosensor\cite{Ciminelli2013}. In his paper of 2002\cite{Vollmer2002}, Vollmer and colleagues measured the resonance-wavelength shift induced  by a layer of proteins on an handcrafted silica microsphere.
Since then, many other authors have elaborated this concept, improving its scalability\cite{DeVos2009,Washburn2010a}, the sensitivity of the resonator\cite{Gaathon2006}, the quality factor\cite{Armani2003,Lee2012}, the integration of the sensors in complex and automated system\cite{Iqbal2010} and the biological functionalization technique for the specific recognition of the target biomolecule\cite{Pasquardini2013}.

We recently demonstrated the possibility to integrate on-chip a monolithic free-standing disk resonator with a vertically-coupled bus waveguide\cite{Ghulinyan2011}. Our further works showed how the vertical coupling architecture is particularly suited to couple integrated high quality-factor (Q) wedge resonators, where the sidewall of the resonator is tilted with respect to the surface of the substrate\cite{Ramiro-Manzano2012}.

The very high Q is a key aspect for the realization of sensors with the highest resolution. In this sense, wedge resonators could be good candidates for the realization of hi-performance integrated sensors. In this paper, we investigate the advantages of these structures from the point of view of their application as refractive index sensors. In the first section, we experimentally characterize the optical sensing parameters of a wedge resonator and compare them to those of a disk resonator with similar dimensions. 
In order to analyze the effects of the shape of the structures under test from a sensor point of view, we pose the basis and the model for their analysis. 
Finally, we solve numerically the optimization problem by means of finite-element-analysis. We find the critical geometrical parameters for achieving best sensing performances and we compare the results to that of the measured samples.

\section{Experimental observations}
\label{sec:experiments}
Wedge resonators are attractive because of their demonstrated feasibility of obtaining on-chip ultra-high quality-factors, which is particularly appealing for many applications\cite{Lee2012}. This class of devices are WGM optical resonators lithographically defined by means of an isotropic etching step. Therefore, their sidewalls are not vertical, and the angle with the substrate can be varied by tuning the etching parameters. 

\begin{figure}[htbp]
\centering
\includegraphics[width=0.9\textwidth]{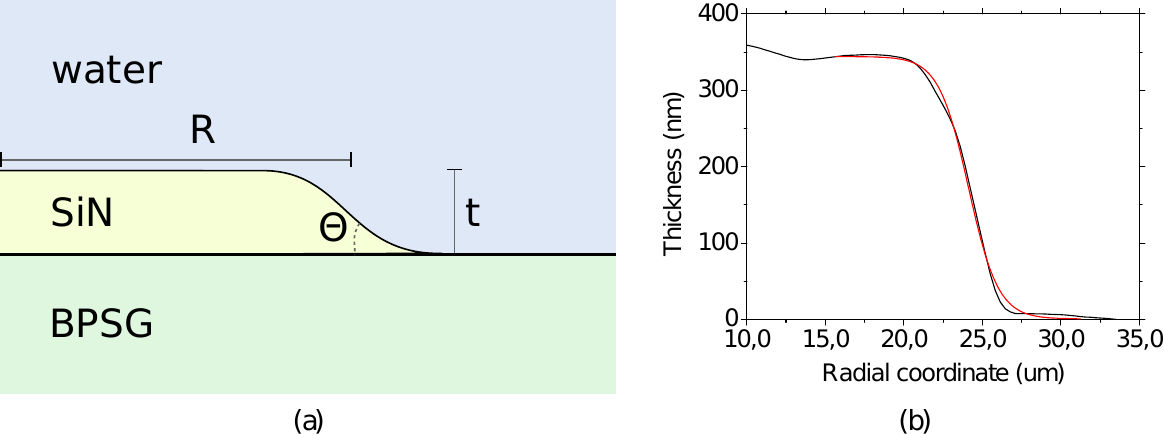}
\caption{(a) Geometrical model of the wedge WGM resonator. The variables are the parameters used in the analysis of this paper. (b) AFM profile of the sidewall of a wedge resonator. The red line is the fitted logistic function \eref{eq:logistic}.}
\label{fig:1}
\end{figure}

In \fref{fig:1}(a) a simple scheme of a wedge WGM resonator is shown, with the definitions of radius ($R$), thickness ($t$) and inclination ($\theta$). AFM images of some of these structures (\fref{fig:1}(b)) show that a trapezoidal shape is a simplicistic model for the real profile of the wedges. In this work we used the logistic function 
\begin{equation}
\label{eq:logistic}
f(x) = t \left[ 1 + \exp(k (x-R)) \right]^{-1}
\end{equation}
to model the sidewalls, where \mbox{$k=4\tan(\theta)/t$} and $x$ is the radial coordinate, with origin in the center of the disk. On our samples this function gives best fits to the profile measurements, as can be seen in \fref{fig:1}(b).

The resonant structures herein analyzed were intended to be operated in the NIR-visible spectrum. Thus the chosen dielectrics are:
\begin{itemize}
\item silicon nitride (SiN$_x$) for the core, for its high refractive index and transparency;
\item borophosphosilicate glass (BPSG) for the lower cladding, to allow for a cost effective fabrication of the monolithic vertical coupling to bus waveguide\cite{Ghulinyan2011};
\item water for the upper cladding (label-free experiments are actually carried out in buffer solutions, which usually have a refractive index very similar to that of water $\pm 1$\%).
\end{itemize}

In order to compare the sensing properties of the commonly used disk resonators (i.e. those with a vertical sidewall) versus wedge resonators, we have run two fabrication processes. These two runs have in common the same PECVD deposition steps and same lithography masks, but they differ in the etching procedure during the resonator definition. The resulting devices are $~350\,$nm thick, with a radius $R=25\,\mu$m and $R=24\,\mu$m for disk and wedge respectively. The disk edge inclination is $\theta\sim 85$\textdegree, while the wedge inclination is $\theta\sim 7$\textdegree. Both resonators are vertically coupled to an integrated silicon oxynitride bus waveguide. Details on the fabrication process are reported in \cite{Ramiro-Manzano2012}. 

Laser light with wavelength close to $1550\,$nm can be coupled in these two structures. Depending on the horizontal displacement of the bus waveguide with respect to the edge of the microresonators the coupling changes and different resonance families can be excited. In this work this parameter has been adjusted to excite the first and second radial quasi-TE family modes. We measured the spectral position and quality factor of these resonances as a function of the refractive index of the liquid in contact with the sensor surface. As explained and motivated in section \ref{sec:theory}, from this data we calculated the bulk sensitivity $S$ and the bulk figure of merit $\mathrm{FOM_{b}}$:
\begin{align}
\label{eq:Sb_simple}
S &= \frac{\partial{\lambda}}{\partial{n}} \\
\mathrm{FOM_{b}} &= \frac{S}{\Gamma} = \frac{Q S}{\lambda}
\end{align}
where $\lambda$, $\Gamma$ and $Q$ are the resonance wavelength, linewidth and quality factor, respectively, and $n$ is the refractive index of the sensing liquid.

During the experiments, the temperature of the samples was controlled with a Peltier element, and a drop of the liquid to be measured was dispensed with an Eppendorf Femtojet.
To change the refractive index of the sensing liquid we prepared several water-glucose solutions, with concentrations spanning from $0\%$ to $0.5\%$w/w, which provide a refractive index variation of up to $6.5\cdot 10^{-4}\,$RIU.
Immediately before and after the dispensing, the spectrum of the resonator has been measured using a swept laser at wavelengths in the range $1530-1560\,$nm. Thereupon, the surface of the sensor was then rinsed with excess of pure water and dryed with a nitrogen flow for the next measurement.

\begin{figure}[htbp] 
\centering
\includegraphics[height=4.8cm]{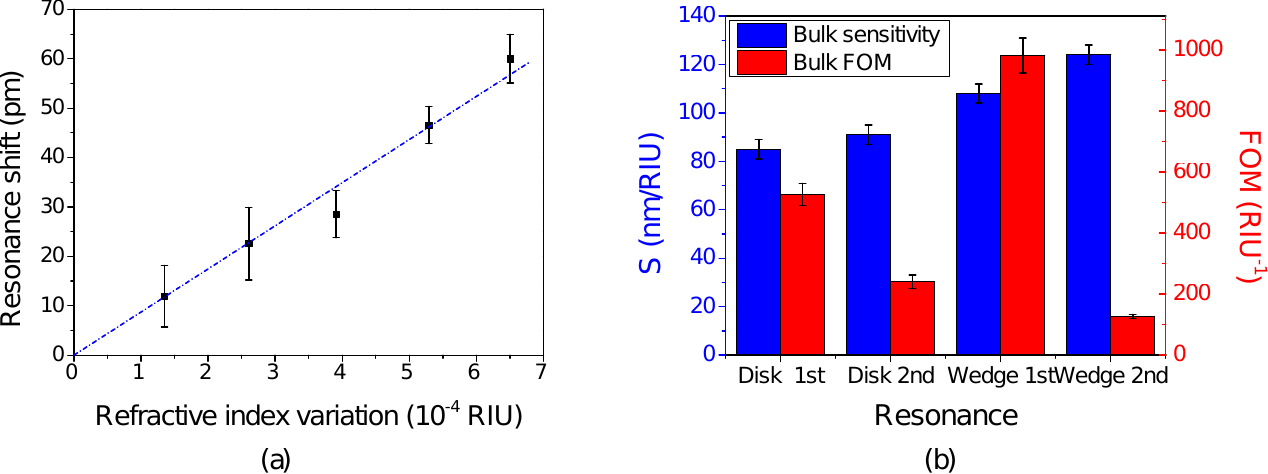}
\caption{(a) Experimental data showing the linear relation between resonance wavelength and bulk refractive index of water-glucose solutions for the first radial TE mode of a disk resonator. (b) Bulk sensitivity and figure of merit measured on the first and second TE radial-family WGMs of disk and wedge resonators. We considered resonances with center wavelengths close to $1540\,$nm.}
\label{fig:exp}
\end{figure}
In \fref{fig:exp}(a) we show an example of the measured resonance shift and we compare the sensing performances of the disk and wedge resonators in \fref{fig:exp}(b). We notice that, compared to the commonly used first radial resonance of the disk resonator ($S = 85\pm4\,$nm/RIU), the bulk sensitivity is enhanced by either using the second radial family ($S = 91\pm4\,$nm/RIU) or reducing the inclination ($S = 108\pm4\,$nm/RIU). This is reasonable, since both choices have the effect of reducing the confinement of the WGM and, therefore, of increasing the intereaction of the mode with the sensing liquid. However, the WGMs of the second radial family exhibit lower quality factors. We measured a quality factor $Q = 4100\pm200$ and $Q = 1580\pm30$ for the second radial families of disk and wedge, respectively, to be compared with $Q = 9600\pm200$ and $Q = 14100\pm300$ for the first radial families. This is the reason why the highest bulk FOM is achieved by the wedge's first radial WGM family. Compared to the disk, which exibits a FOM$=530\pm40\,$RIU$^{-1}$, the wedge performs almost twice better, with a FOM$=980\pm60\,$RIU$^{-1}$.

In order to understand such differences, we decided to analyze by numerical simulation the influence of the geometry of the wedge resoantor on the device performance.

\section{Problem analysis and sensor design}
\label{sec:theory}
The problem of the optimization of the sensor's parameters can be adressed by different strategies. For example, in a recent paper the author's aim was to maximize the sensitivity to analyte while minimizing the thermal sensitivity\cite{Delezoide2014}. A different approach is used in \cite{Foreman2014}, where the optimal parameters for minimizing the analyte detection limit are extensively studied starting from the measurement of the amplitude, position and linewidth of the resonance. In this paper, we use an approach similar to that of \cite{Ciminelli2014}, whereby we analyze both the sensitivity and quality factor of the supported resonant modes as a function of geometrical parameters of the resonator. Thus, to optimize the sensor performances, one should ideally know the sensitivity and the quality factor of the WGMs supported by the possible resonator configurations. 

The sensitivity \eref{eq:Sb_simple} can be calculated from the optical mode profile $\vec{E}(\vec{r})$. By using a perturbative approach\cite{Johnson2002,Teraoka2003}, it can be shown that the wavelength sensitivity $S_{V,\,i}$ of the $i$-th resonant mode to a variation of the refractive index in the domain $V$ can be expressed as
\begin{equation}
S_{V,\,i} = \frac{\partial{\lambda_i}}{\partial{n_V}} = \frac{\lambda_i}{n_V} \frac{\iiint_V \epsilon(\vec{r}) |\vec{E}_i(\vec{r})|^2 d^3\vec{r} }{\iiint_{all}  \epsilon(\vec{r}) |\vec{E}_i(\vec{r})|^2 d^3\vec{r}}\;\; .
\label{eq:sensitivity_bulk}
\end{equation}
Here the two integrals are taken in the sensing volume $V$ and in the whole model, respectively, and $n_V$ is the unperturbed refractive index of the homogeneous medium contained in $V$. In the same way, in both integrals the dielectric constant $\epsilon(\vec{r})$ and the electric field $\vec{E}_i(\vec{r})$ are meant to be unperturbed, i.e. when no analyte is present. As it can be seen, the sensitivity density is proportional to the electric field energy density $\frac{1}{2}\epsilon|\vec{E}|^2$. Due to the complex gemetrical structure, finite element method (FEM) modeling of the resonator has been used to obtain the mode profiles $\vec{E}_i(\vec{r})$. 
We assume that the mode profile in the cavity is not significantly altered by the presence of the waveguide. Thus, the model used in this work can ignore the presence of the bus waveguide. The advantage of this choice is that the cylindrical symmetry of the structure is not broken, and a simple and fast axisymmetric 2.5D simulation can be performed. 

When light is confined in a waveguide or in a cavity, it is quite common to obtain evanescent tails which extend from the surface of the structure for tens to hundreds of nanometers\cite{Rigo2013,Aparicio2014}. When used as label-free sensors, the volume of interaction between the analyte and the WGM evanescent field is very small, and mainly constrained by the thickness of the layer of captured analyte. In the case of nanometric-sized molecules (like proteins) this means that most of the evanescent tail is unperturbed and not used to produce a signal. For a fair comparison  between different structures, it is very helpful to introduce the superficial sensitivity $\sigma_{A,\,i}$ defined as
\begin{equation}
\sigma_{A,\,i} = \frac{\lambda_i}{n_V} \frac{\iint_A \epsilon(\vec{r}) |\vec{E}_i(\vec{r})|^2 d^2\vec{r} }{\iiint_{all}  \epsilon(\vec{r}) |\vec{E}_i(\vec{r})|^2 d^3\vec{r}} = \frac{\partial^2{\lambda_i}}{\partial{t_V}\partial{n_V}} \approx \frac{S_{V,\,i}}{t_V} 
\label{eq:sensitivity_sup}
\end{equation}
where $A$ is the area of the bottom surface of a thin layer of volume $V$ and thickness $t_V$. The approximation in \eref{eq:sensitivity_sup} is valid for layers $V$ much thinner than the extension of the evanescent field ($t_V \lessapprox 20\,$nm). Please note the double partial derivative in \eref{eq:sensitivity_sup}: this is different from the definition usually adopted ${\partial{\lambda_i}}/{\partial{t_\ell}}$ (see for example \cite{Ciminelli2014}) which is not independent from the refractive index of the analyte assumed in the volume $V$. The definition that we adopted is independent from the refractive index $n_V$ and can be used for direct comparison between different structures or models.

Calculating the quality factor is a more subtle issue. By taking into account the losses of the structure, the hermiticity of the eigenvalue problem is broken and the solutions are complex-valued. From the complex effective refractive index or from the complex eigenfrequency $\omega$ of the resonant mode, the quality factor can be calculated as
\begin{equation}
Q_i = \frac{\Re(\omega_i)}{2 \Im(\omega_i)}
\label{eq:quality}
\end{equation}
where $i$ is the subscript for the $i$-th resonant mode and $\Re(\omega_i)$ and $\Im(\omega_i)$ are the real and imaginary part of $\omega_i$, respectively. In our model we account for radiative losses, absorption losses and coupling losses. The firsts are modelled implementing perfect matched layers (PML) at the boundaries of the simulation\cite{Cheema2013}, the seconds are added in both water and core material as imaginary part of the refractive index, while the last are simply considered by dividing by two the quality factor (critical coupling regime). We neglected other sources of losses according to our previous (unpublished) measurements on similar resonators.

To evaluate the bulk and superficial sensing performances and choose the optimal set of parameters, we use two figures-of-merit (FOM) calculated from the sensitivities and quality factors of every resonance:
\begin{subequations}
\begin{align}
\mathrm{FOM_{b}} &= \frac{S_{V,\,i}}{\Gamma_i} = \frac{Q_i S_{V,\,i}}{\lambda_i}
\label{eq:FOM_bulk}
\\
\mathrm{FOM_{s}} &= \frac{\sigma_{A,\,i}}{\Gamma_i} = \frac{Q_i \sigma_{A,\,i}}{\lambda_i}\;\; .
\label{eq:FOM_layer}
\end{align}
\end{subequations}
From now on, all the subscripts $i$ used to label the different modes are dropped for simplicity of reading and, if not differently specified, $V$ and $A$ are meant to be the whole sensing volume/area above the parts of the sensor exposed to analytes.

This definition of figure of merit is equivalent to that used in surface plasmon resonance (SPR) label-free sensors\cite{Shen2013}. With this figure of merit it is possible to directly compare the performances of sensors exploiting different techniques. A table which reviews the FOM of WGM resonators, photonic crystals (PhC) and SPR sensors is reported in \cite{Huang2014}. In addition, we also introduce the superficial figure of merit $\mathrm{FOM_{s}}$, which aims to compare the performances of surface sensors, used for example in label-free detection.

If the resolution of the resonance wavelength measurement is limited by the resonance linewidth itself and not from other experimental parameters, one can estimate the achievable limit of detection (LOD) as
\begin{equation}
\mathrm{LOD_{b/s}} = \frac{1}{\eta} \frac{1}{\mathrm{FOM_{b/s}}}
\label{eq:LOD}
\end{equation}
where $\eta$ accounts for an enhanced resolution given by a proper resonance fitting procedure. This parameter $\eta$ depends on experimental details and usually lays in the range $10\sim 100$\cite{Vollmer2012}.

\section{Wedge geometry optimization}
Using a commercial FEM solver, we have modelled the geometry shown in \fref{fig:1} and we solved for the electric field's eigenfunctions supported by the structure. We varied the value of the wedge inclination from $\theta=2$\textdegree $\;$ to $\theta\sim 90$ \textdegree, and the wedge thickness from $t=200\,$nm to $t=500\,$nm. For every solution we calculated the bulk and superficial sensitivity, $S$ and $\sigma$,  the quality factor, $Q$, and the two figures of merit, $\mathrm{FOM_{b/s}}$. The other geometrical parameters were kept fixed, and their values are reported in \tref{tab:1}.
\begin{table}[htbp]
\begin{center}
\caption{Parameters used in the optimization study.}
\label{tab:1}
\begin{tabular}{|c|c|l|}
\hline 
Parameter & Value & Description \\
\hline
$R$ & $24\,\mu$m & resonator radius \\
$\lambda$ & $\sim 1540\,$nm & resonance wavelength \\
$n_{\mathrm{water}}$ & $1.32$ & water refractive index (real part) \\
$k_{\mathrm{water}}$ & $1\cdot 10^{-4}$ & water refractive index (imaginary part) \\
$n_{\mathrm{SiN}}$ & $1.99$ & SiN$_x$ refractive index (real part) \\
$k_{\mathrm{SiN}}$ & $5\cdot 10^{-5}$ & SiN$_x$ refractive index (imaginary part) \\
$n_{\mathrm{BPSG}}$ & $1.46$ & BPSG refractive index (real part) \\
$k_{\mathrm{BPSG}}$ & $0$ & BPSG refractive index (imaginary part)\\
$t_V$ & $5\,$nm & auxiliary layer thickness (for $\sigma$) \\
\hline
\end{tabular}
\end{center}
\end{table}

\begin{figure}[htbp] 
\centering
\includegraphics[width=\textwidth]{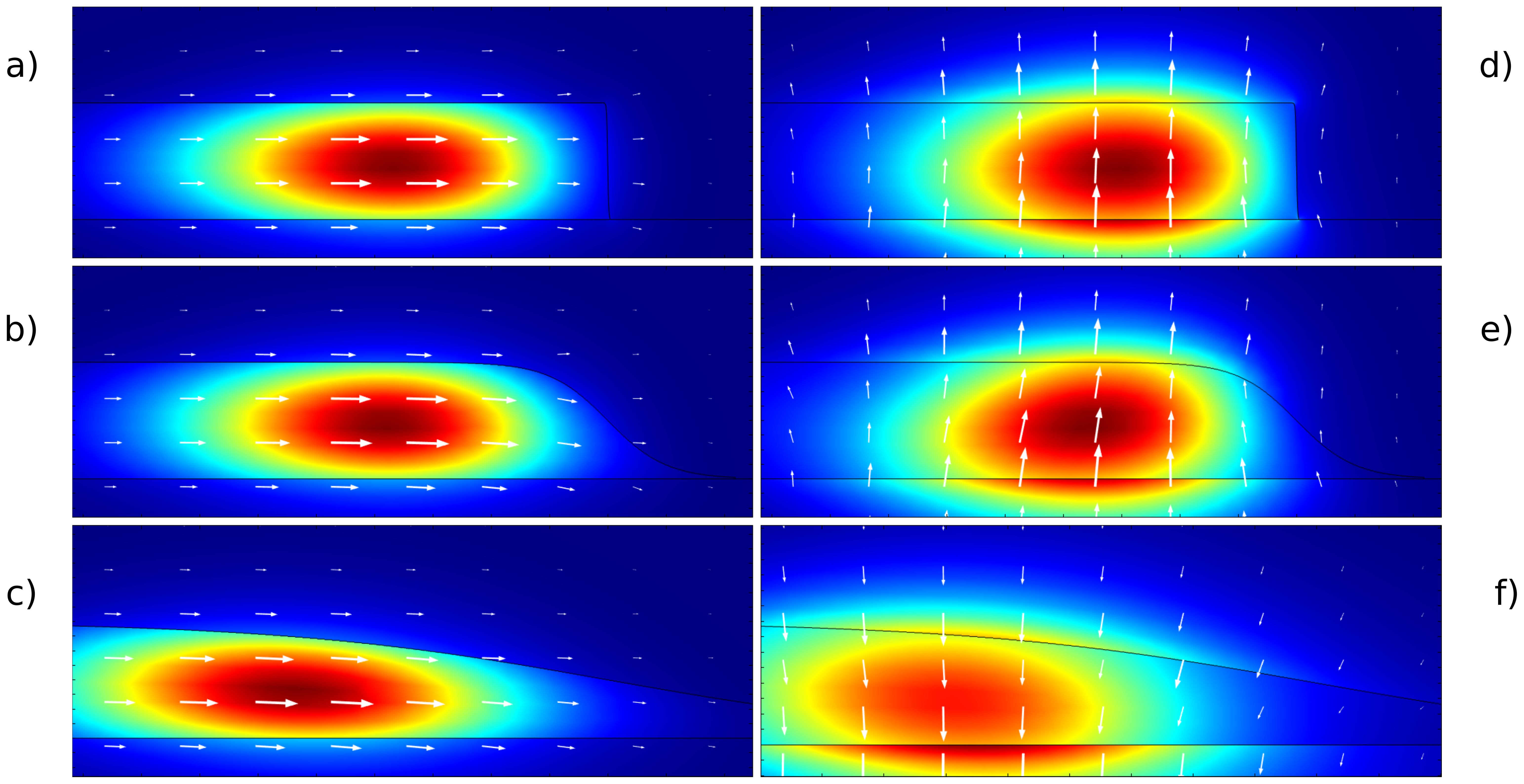}
\caption{Electric field energy density for the first radial TE (a,b,c) and TM (d,e,f) modes for wedge angles of 89 (a,d), 45 (b,e) and 10 (c,f) degrees. The white arrows depict the orientation of the electric field. The resonator's thickness is $t=400\,$nm.}
\label{fig:modes}
\end{figure}
In \fref{fig:modes} we show the first radial TE and TM modes for three different wedge angles and same wedge thickness $t=400\,$nm. The difference in the confinement and distribution of the electric field is evident, particularly when comparing the two polarizations. Slightly more difficult to notice, but still very interesting, is the fact that the mode reaches its highest confinement for intermediate wedge angles (roughly between $30$\textdegree$\,$ and $60$\textdegree, depending on the other parameters). For greater angles, the field extends mainly from the top surface of the resonator, while for smaller angles the field can sense the area above the external sidewall. 

\begin{figure}[htbp]
\centering
\includegraphics[width=\textwidth]{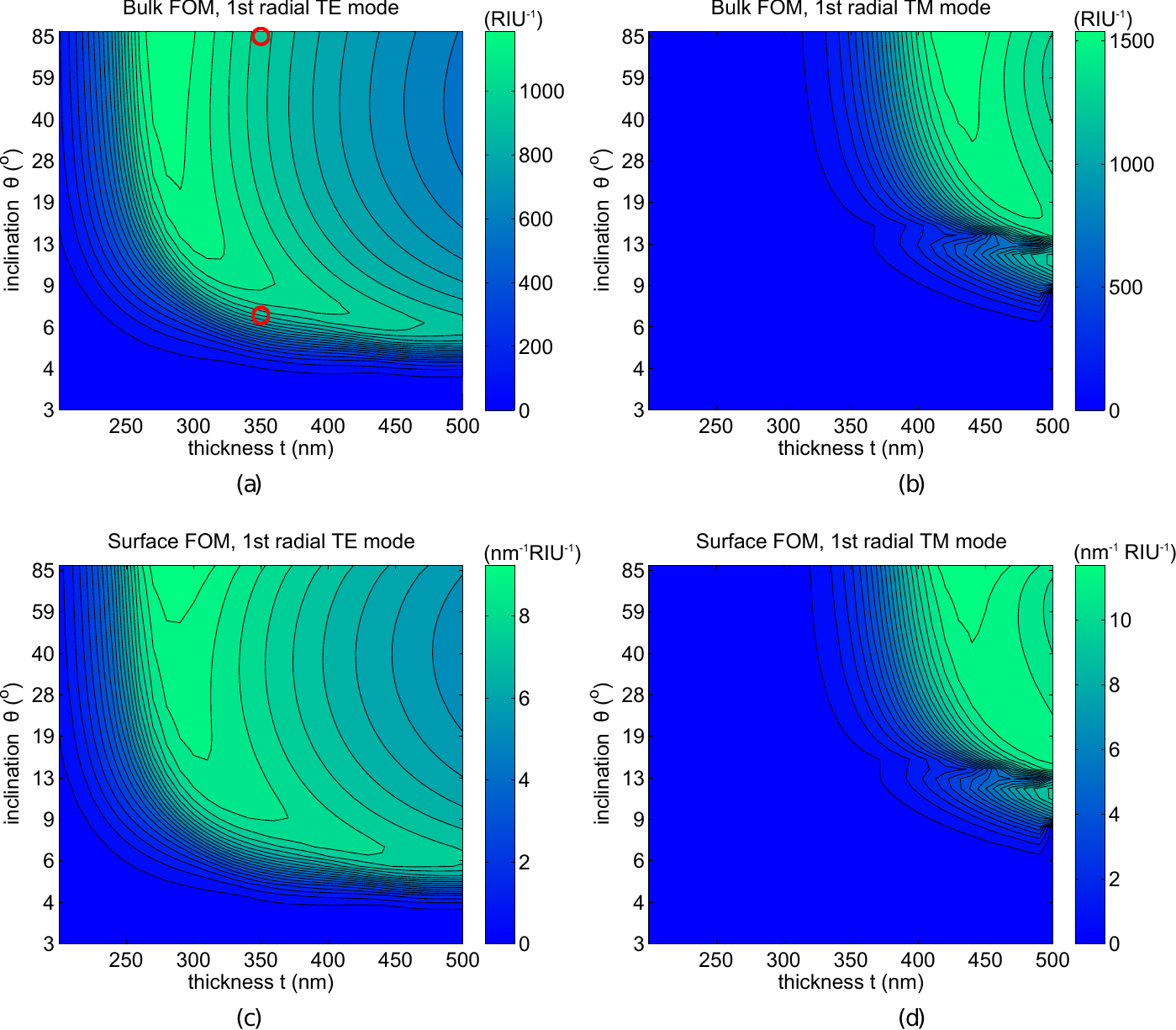}
\caption{Optimization analysis performed on the first radial TE (a,c) and TM (b,d) modes, as a function of the wedge inclination $\theta$ and thickness $t$. The plots report the bulk (a,b) and superficial (c,d) figure of merit $\mathrm{FOM_{b/s}}$. The red circles in (a) mark the properties of the samples characterized in  section \ref{sec:experiments}.}
\label{fig:2}
\end{figure}

In \fref{fig:2} we report the 2D contour-plot summarising the optimization analysis of the bulk and surface FOM for the first radial TE and TM modes.

For what concerns the bulk sensing (sensor used as refractometer for liquids), it is interesting to notice that a wide set of parameters can be used to achieve good performances. Once that the thickness has been optimized (i.e. $t \sim 280\,$nm for TE polarization and $t \sim 430\,$nm for TM polarization), the value of the inclination has little impact on the bulk FOM, at least for $\theta \gtrsim 40$\textdegree. 
In this configuration we get $\mathrm{FOM_b}\gtrsim 1200/{\mathrm{RIU}}$ for TE polarization and $\mathrm{FOM_b}\gtrsim 1600/{\mathrm{RIU}}$ for TM polarization. 
Conversely, the choice of the optimal set of parameters for a sensor used for surface sensing (e.g. a label-free biosensor) is much more strict. The best FOM is achieved only with a vertical wedge angle (i.e. a disk resonator) both for TE ($\mathrm{FOM_s}\gtrsim 9.2/{\mathrm{nm\;RIU}}$) and TM polarization ($\mathrm{FOM_s}\gtrsim 11.7/{\mathrm{nm\;RIU}}$). 
Interesting is the fact that the TM polarization can provide higher FOM if the resonator is properly optimized. For bulk sensing the enhancement can be higher than $30\%$, while for surface sensing the enhancement can exceed $20\%$. 

Regarding the experimental results of section \ref{sec:experiments}, the first observation is that the thicknesses of the wedge and disk resonators were not optimal for sensing applications.
For a resonator with $t=350\,$nm, the modes with the highest FOM (both bulk and surface) are the first two radial TE modes, which was also observed experimentally. In \fref{fig:2}(a), the red circles label the parameters of the measured sensors on the bulk FOM map. 

\begin{figure}[htbp]
\centering
\includegraphics[width=9cm]{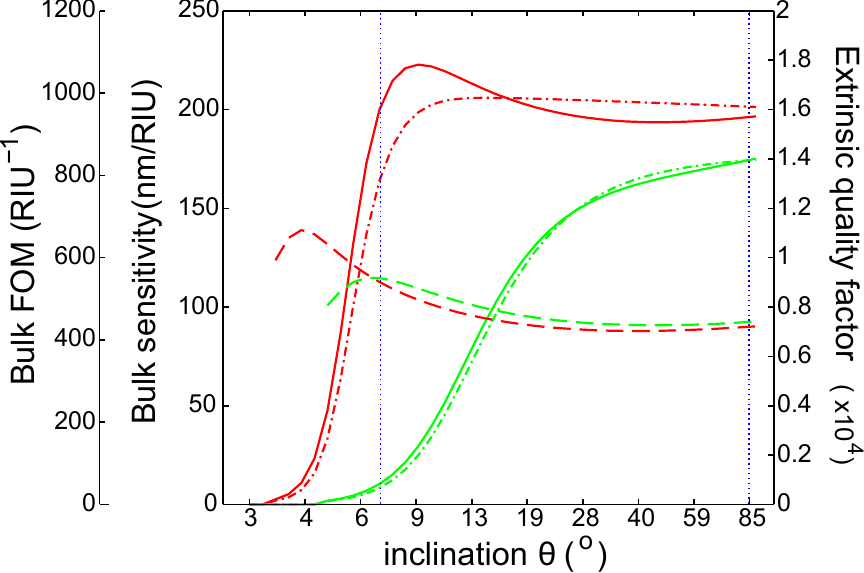}
\caption{Quality factor (dashed-dot line), bulk sensitivity (dashed line) and bulk FOM (continuous line) for the first (red) and second (green) TE radial modes. The vertical blue lines are markers for the parameters of the fabricated sensors. The resonator thickness is $t=350\,$nm.}
\label{fig:3}
\end{figure}
To compare the experimental results with the simulations, \fref{fig:3} reports the bulk FOM together with the bulk sensitivity and quality factor of the two aforementioned TE modes.
The figure qualitatively confirms the experimental results: for this thickness the wedge resonator can provide a higher bulk figure of merit with respect to a disk resonator. Similarly, the bulk sensitivity is higher for $\theta=7$\textdegree$\;$ than for $\theta=85$\textdegree$\;$ and even higher for the second mode in both structures. 
Finally, a note should be done about the accuracy of the quality factor in the simulations: in our model we didn't take into account possible variations in the scattering losses due to the roughness of the resonator. In particular, the two compared structures were defined with two different etching processes, which are known to lead to different final roughnesses. This can explain why the real quality factor, as well as the bulk FOM, are actually lower than expected in the case $\theta=85$\textdegree. The same does not apply to the case of the bulk sensitivity, since $S_b$ does not depend on the roughness of the structure.

The results of our analysis clearly show that the control on the resonator thickness is the most important parameter to achieve optimal sensing. The use of vertical coupling is thus necessary to have this freedom in the fabrication without imposing limitations on the geometry of the bus waveguide\cite{Ghulinyan2011}. This technique, in fact, relies on two deposition and lithography steps, where the thicnkess of both deposited layers can be changed independently according to the results of the optimization analysis.

\section*{Conclusion}
In this work we tested the effectiveness of vertically-coupled integrated wedge resonators for refractometric sensing applications, with a particular focus to label-free biosensing. These devices are very appealing because of the very high quality factor that they can exhibit. Nevertheless, the quality factor is not the only concern when analyzing the device performances. The figures of merit of interest for sensing applications are also affected by the field overlap with the analyte.

We experimentally verified that under certain conditions a wedge resonator can perform better compared to a disk resonator with similar dimensions. At the same time, however, we also showed, by means of numerical analysis, that optimal design parameters can be achieved by changing the thickness of a normal disk resonator (inclination $\theta \sim 90$\textdegree).
We also showed that the TM-polarization modes can lead to higher performances with respect to TE modes, provided an adequate resonator thickness is used to reduce the radiation losses.

In the modelling we used two figures of merit (bulk and surface) that permit an easy and fair comparison between sensors with different geometries, even between sensors relying on different physical principles (e.g. SPR or PhC). The bulk FOM shows that our structures perform at least one order of magnitude better than SPR devices for bulk refractometric sensing, which are in the order of $\mathrm{FOM_b}\approx100/\mathrm{RIU}$ \cite{Huang2014}. In addition, we introduced the surface FOM, that can be used to compare different structures with special focus on surface sensing (like the label-free method).

To summarize, the use of wedge resonators in place of conventional disk resonators is not suggested, at least for resonators similar to the ones described by \tref{tab:1}. The isotropic etching step is still suggested to obtain better surface quality, but cares have to be taken in order to keep the inclination $	\theta\gtrsim60\,$\textdegree. The analysis here presented shows that the optimization of the resonator thickness is critical to achieve the best performances. In this regard, the use of vertical coupling has emerged to be particularly appealing: not only it avoids any interaction with analyte in the coupling region during sensing, but in addition it gives control on the resonator thickness and geometry without posing limitation to the waveguide. 

\section*{Acknowledgments}
This works was supported by the FP7 EU project "Symphony" and by FU-PAT through the "NAoMI" project. The authors thanks dr. Romain Guider for the fruitful discussions.

\end{document}